\documentclass[twocolumn,showpacs,preprintnumbers,amsmath,amssymb]{revtex4}
\usepackage{graphicx}
\usepackage{dcolumn}
\usepackage{bm}

\begin{document}
\def\d{{\rm d}}
\def\Epos{E_{\rm pos}}
\def\ap{\approx}
\def\eff{{\rm eff}}
\def\L{{\cal L}}
\newcommand{\vev}[1]{\langle {#1}\rangle}
\newcommand{\CL}   {C.L.}
\newcommand{\dof}  {d.o.f.}
\newcommand{\eVq}  {\text{eV}^2}
\newcommand{\Sol}  {\textsc{sol}}
\newcommand{\SlKm} {\textsc{sol+kam}}
\newcommand{\Atm}  {\textsc{atm}}
\newcommand{\Chooz}{\textsc{chooz}}
\newcommand{\Dms}  {\Delta m^2_\Sol}
\newcommand{\Dma}  {\Delta m^2_\Atm}
\newcommand{\Dcq}  {\Delta\chi^2}
\newcommand{\nbb}{$\beta\beta_{0\nu}$ }
\def\VEV#1{\left\langle #1\right\rangle}
\let\vev\VEV
\def\e6{E(6)}
\def\10{SO(10)}
\def\21{SU(2) $\otimes$ U(1) }
\def\321{$\mathrm{SU(3) \otimes SU(2) \otimes U(1)}$ }
\def\lr{SU(2)$_L \otimes$ SU(2)$_R \otimes$ U(1)}
\def\422{SU(4) $\otimes$ SU(2) $\otimes$ SU(2)}
\newcommand{\AHEP}{%
  AHEP Group, Instituto de F\'{\i}sica Corpuscular --
  C.S.I.C./Universitat de Val{\`e}ncia \\
  Edificio Institutos de Paterna, Apt 22085, E--46071 Val{\`e}ncia, Spain}
\newcommand{\Tehran}{%
Institute for Studies in Theoretical Physics and Mathematics
(IPM), P.O. Box 19395-5531, Tehran, Iran}
\def\roughly#1{\mathrel{\raise.3ex\hbox{$#1$\kern-.75em
      \lower1ex\hbox{$\sim$}}}} \def\lsim{\roughly<}
\def\gsim{\roughly>}
\def\ltap{\raisebox{-.4ex}{\rlap{$\sim$}} \raisebox{.4ex}{$<$}}
\def\gtap{\raisebox{-.4ex}{\rlap{$\sim$}} \raisebox{.4ex}{$>$}}
\def\lsim{\raise0.3ex\hbox{$\;<$\kern-0.75em\raise-1.1ex\hbox{$\sim\;$}}}
\def\gsim{\raise0.3ex\hbox{$\;>$\kern-0.75em\raise-1.1ex\hbox{$\sim\;$}}}

\preprint{
 IFIC/05-44\cr IPM/P-2005/068}


\title{R-parity violation assisted thermal leptogenesis in the
   seesaw mechanism}

\date{\today}
\author{Y. Farzan}\email{yasaman@theory.ipm.ac.ir}
\affiliation{\Tehran}
\author{J. W. F. Valle}\email{valle@ific.uv.es}
\affiliation{\AHEP}
\begin{abstract}

  Successful leptogenesis within the simplest type I supersymmetric
  seesaw mechanism requires the lightest of the three right-handed
  neutrino supermultiplets to be heavier than $\sim10^9$~GeV.  Thermal
  production of such (s)neutrinos requires very high reheating
  temperatures which result in an overproduction of gravitinos with
  catastrophic consequences for the evolution of the universe.  In
  this letter, we let R-parity be violated through a $\lambda_i \hat{N}_i
  \hat{H}_u \hat{H}_d$ term in the superpotential, where $\hat{N}_i$ are right-handed
  neutrino supermultiplets.  We show that in the presence of this
  term, the produced lepton-antilepton asymmetry can be enhanced.  As
  a result, even for $\hat{N}_1$ masses as low as $10^6$~GeV or less, we can
  obtain the observed baryon asymmetry of the universe without
  gravitino overproduction.

\end{abstract}
\date{\today}
\maketitle  The recent neutrino data
\cite{fukuda:1998mi,ahmad:2002jz,eguchi:2002dm} shows that
neutrinos are massive~\cite{Maltoni:2004ei}. One of the most
popular ways to generate tiny nonzero neutrino masses is  the
seesaw mechanism~\cite{Minkowski:1977sc,schechter:1980gr}, which
adds three right-handed neutrinos to the standard model with very
heavy masses, $M_3>M_2>M_1\gg m_{weak}$.

 The seesaw mechanism also provides us with a framework
to obtain the observed baryon-antibaryon asymmetry of the universe
through a process called leptogenesis \cite{Fukugita:1986hr}.
However, in the context of supersymmetry, this process suffers
from a phenomenon called gravitino
overproduction~\cite{Coughlan:1983ci}: If we assume that lightest
right-handed (s)neutrinos are thermally produced in the early
universe, the reheating temperature ($T_R$) should be higher than
$\sim 10^{9}$ GeV \cite{Buchmuller:2004nz,Giudice:2003jh}. The
high reheating temperature can lead to the overproduction of
gravitinos which has catastrophic consequences for the evolution
of the universe. The upper bound on $T_R$ from gravitino
overproduction considerations depends on the details of model.  If
the gravitino has hadronic decay modes, we expect $T_R< 10^{6-7}$
GeV \cite{Kawasaki:2004qu}. In the literature, a variety of
solutions for this problem has been suggested
\cite{Pilaftsis:1997jf,Ellis:2002eh,Bolz:1998ek,Ibe:2004tg,Abada:2003rh,Dent:2005gx}.

In this letter, we suggest an alternative solution based on the
R-parity violation. The produced lepton-antilepton asymmetry can be in
the expected range, even for masses of the lightest of the
right-handed (s)neutrinos lower than $10^6$~GeV, avoiding gravitino
overproduction.

First recall that in the simplest type I supersymmetric seesaw
mechanism the superpotential is given by
\begin{eqnarray}
\label{eq:usual}
W&=&\sum_{i,j}\epsilon_{\alpha \beta}(Y_\nu)_{ij}
\hat{N}_i \hat{L}_j^\alpha \hat{H}_u^\beta +\frac{1}{2}\sum_{ij}M_{ij} \hat{N}_i
\hat{N}_j
\end{eqnarray}
where $\hat{L}_j$ is the superfield associated with the left-handed
lepton doublet $(\hat{\nu}_{j} , \hat{l}_{j})$ and $\hat{H}_u$ is the
Higgs doublet that gives mass to the up quark.  The first term is the
familiar Yukawa coupling and the second is the Majorana mass term of
right-handed neutrinos.

Relaxing R-parity conservation, we can add the following term to the
superpotential
\begin{equation}
\label{eq:new} W_{\rm RPV}= \sum_i \epsilon_{\alpha
\beta}\lambda_i \hat{N}_i \hat{H}_d^\alpha \hat{H}_u^\beta
\end{equation}
where $\hat{H}_d$ is the Higgs doublet that gives mass to the down
quark.  The existence of this R--Parity violating (RPV) term has
recently been advocated to solve the $\mu$
problem~\cite{Lopez-Fogliani:2005yw}.  In our case its
contribution to generating the $\mu$ term is negligible because
$\tilde{N}_i$, being super-heavy, do not acquire sizeable vacuum
expectation values. However this term will play a key role in
making thermal seesaw leptogenesis viable.

Note that R--Parity violation in supersymmetry has been advocated as
an attractive origin for neutrino masses, alternative to the
supersymmetric seesaw~\cite{Hirsch:2004he}. Neutrino masses are
typically hierarchical, with the atmospheric scale arising at tree
level and the solar one calculable as radiative
corrections~\cite{Hirsch:2000ef}. However here we propose that
neutrinos acquire masses {\it a la seesaw} and that, although RPV is
necessary to produce the observed baryon asymmetry of the universe, it
is not the dominant source of neutrino masses.

Without loss of generality, we can rotate and rephase the fields
to make the mass matrix $M_{ij}$ real diagonal. In this basis, the
elements of $Y_\nu$ and $\lambda$ can in general be complex.
Introduction of the coupling $\lambda_i$ adds three extra
CP-violating phases to the theory, as we shall see, with
consequences for the baryon asymmetry of the universe.

Let us define the following asymmetries
\begin{eqnarray}\epsilon_{N_1} &=& -\sum_i\left[\frac{\Gamma(N_1
\to \bar{l_i}\bar{H}_u) - \Gamma(N_1 \to l_iH_u)
}{\Gamma_{\rm tot}(N_1)/2}\right.\cr &+&\left. \frac{\Gamma(N_1 \to
\bar{\tilde{l_i}}\bar{\tilde{H}}_u) - \Gamma(N_1 \to
\tilde{l}_i\tilde{H}_u) }{\Gamma_{\rm tot}(N_1)/2}\right]
\end{eqnarray}
and
 \begin{eqnarray}\epsilon_{\tilde{N}_1} &=&
-\sum_i\left[\frac{\Gamma(\tilde{N}_1 \to
\bar{l_i}\bar{\tilde{H}}_u) - \Gamma(\tilde{N}^*_1 \to
l_i\tilde{H}_u) }{\Gamma_{\rm tot}(\tilde{N}_1)/2}\right.\cr &+&\left.
\frac{\Gamma(\tilde{N}_1^* \to \bar{\tilde{l_i}}\bar{H}_u) -
\Gamma(\tilde{N}_1 \to \tilde{l}_i{H}_u)
}{\Gamma_{\rm tot}(\tilde{N}_1)/2}\right]
\end{eqnarray}
where $N_1$ and $\tilde{N}_1$ are respectively the lightest
right-handed neutrino and sneutrino and $\Gamma_{\rm tot}(N_1)$ and
$\Gamma_{\rm tot}(\tilde{N}_1)$ are their total decay rates. We expect
the produced lepton-antilepton asymmetry to be proportional to
$\epsilon \equiv \epsilon_{N_1}+\epsilon_{\tilde{N}_1}$.
\vspace{0.3cm}

 In the following, we show that the R-parity violating term that we
 have introduced gives a new contribution to $\epsilon_{N_1}$ and
 $\epsilon_{\tilde{N_1}}$ which for certain range of parameters can
 enhance the effect. We show that as a result of this enhancement,
 even for $M_1$ as low as $10^6$ GeV, we can have successful
 leptogenesis and simultaneously generate tiny masses for neutrinos
 [i.e., in the simplest type-I seesaw $(Y_\nu)_{ij} \lsim \sqrt{
   (\Delta m_{atm}^2)^{1/2} M_i /(v^2 \sin^2 \beta)}\sim 10^{-5}
 \sqrt{M_i/(10^6~{\rm GeV})}$, where $v=245$ GeV and
 $\beta=\arctan(\langle H_u \rangle/ \langle H_d \rangle)]$.
 Therefore, thermal production of $N_1$ and $\tilde{N}_1^{(*)}$ does
 not need too high reheating temperature and the universe would not
 encounter gravitino overproduction.
 
 Fig. 1 shows the structure of the diagrams contributing to
 $\epsilon_{N_1}$ and $\epsilon_{\tilde{N}_1}$.  Each line
 collectively represents the bosonic, fermionic or auxiliary component
 of the indicated superfield. The vertices marked with dots are Yukawa
 vertices while those marked with $\otimes$ are the new R-parity
 violating vertices given by $\lambda_i$.  Each line can be either
 bosonic or fermionic when appropriate.  Reversing the arrows we reach
 the diagrams that produce antileptons instead of leptons.
\begin{figure}[htbp]
  \centering
  \includegraphics[bb=33 429 417 756,clip=true,width=\linewidth]{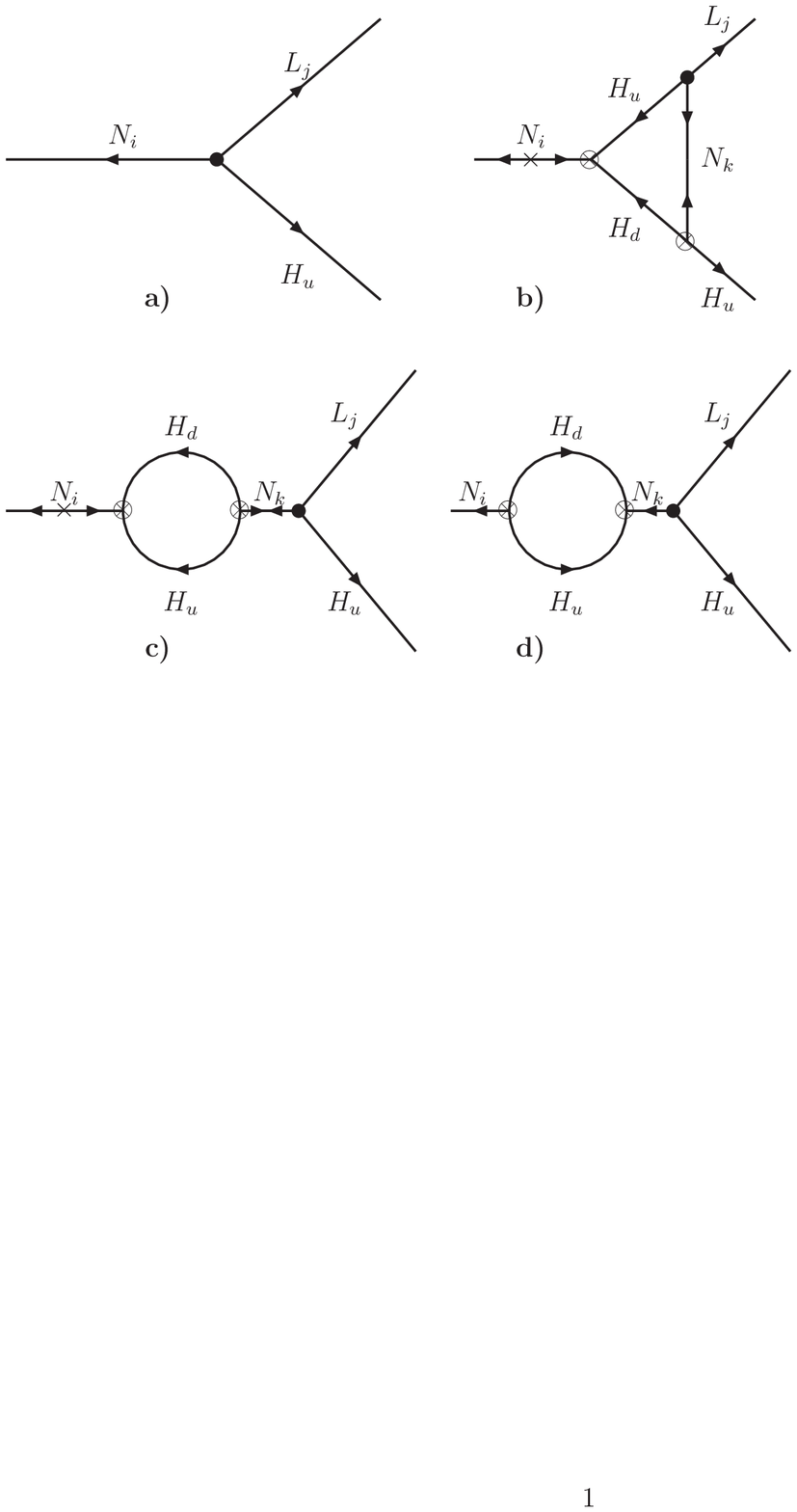}
  \caption{Diagrams contributing to lepton-antilepton asymmetry.
    Vertices marked with dots and $\otimes$ denote Yukawa ($Y_\nu$)
    and R-parity violating ($\lambda$) couplings, respectively.  The
    $\times$ indicates mass insertion.  }
  \label{fig:diagrams-lep}
\end{figure}

Notice that both in the vertex-type diagram (b) and
wave-function-type diagram (c) if we replace $H_d$ by $L_k$, we
will arrive at the familiar diagrams of the standard leptogensis
scenario, see e.g. \cite{Covi:1996wh}. Diagrams (b) and (c)
involve a $\Delta L=2$ Majorana mass insertion in the internal
$N_k$ line ($N_k^T C N_k$ or  $F_{N_k} \tilde{N}_k$).
 There is, however, a new diagram, (d), that
does not have a counterpart in the standard R--parity conserving
case. Notice that, in contrast to the $N_k$ propagator in diagram
(c), the one appearing in diagram (d) is lepton number conserving.

To leading order, we have
\begin{equation}
  \label{eq:vv}
\Gamma_{\rm tot}(N_i)=\Gamma_{\rm tot}(\tilde{N}_i)={(Y_\nu
Y_\nu^\dagger)_{ii} + |\lambda_i|^2 \over 4\pi}M_i
\end{equation}
so that 
\begin{equation}
\epsilon = \frac{1}{2\pi} \sum_{k\ne 1} \left[ [ g(\frac{M_k^2}{M_1^2})+ {2
\frac{M_k}{M_1} \over \frac{M_k^2}{M_1^2}-1}]{\cal I}_{k1} -{2
{\cal J}_{k1}\over \frac{M_k^2}{M_1^2}-1}\right],
\end{equation}
where $g(x)=\sqrt{x} {\rm ln}[(1+x)/x],$
\begin{equation}
  \label{eq:vv}
{\cal I}_{k1}={\sum_j{\rm Im}[(Y_\nu^*)_{1j}\lambda_1^* \lambda_k
(Y_\nu)_{kj}] \over (Y_\nu Y_\nu^\dagger)_{11}+ |\lambda_1|^2}
\end{equation}
and
\begin{equation}
{\cal J}_{k1}={\sum_j{\rm Im}[(Y_\nu^*)_{1j}\lambda_1
\lambda_k^* (Y_\nu)_{kj}] \over (Y_\nu Y_\nu^\dagger)_{11}+
|\lambda_1|^2}.
\end{equation}
Notice that the term proportional to ${\cal J}_{k1}$ comes from the
interference of the tree-level diagram with diagram (d).

Let us suppose $M_1<10^6$ GeV so that thermal production of $N_1$ and
$\tilde{N}_1$ in the early universe can take place without requiring
problematic very high reheating temperatures \cite{Kawasaki:2004qu}.
Moreover let us suppose $M_2$ is not much heavier; $M_2^2/M_1^2\sim
10$. (Since the mechanism we are describing is effective with two
right-handed neutrinos, here we only concentrate on $N_1$ and $N_2$
dropping $N_3$ from the discussion. In principle $N_3$ can play a
similar role as $N_2$.)  For these values of $M_i$, to suppress the
masses of left-handed neutrinos in the simplest type-I seesaw down to
$\sqrt{\Delta m_{atm}^2}$, the Yukawa couplings have to be very tiny $
(Y_\nu)_{ij}\stackrel{<}{\sim} 10^{-5} \sqrt{M_i/10^6 \ {\rm GeV}}$,
similar to that of the electron in the Standard Model. In order for
$N_1$ and $\tilde{N}_1$ to decay out of equilibrium (i.e.,
$\Gamma_{\rm tot}(N_1)=\Gamma_{\rm tot}(\tilde{N}_1)< H|_{T=M_1}$,
where $H$ is the Hubble expansion rate) $\lambda_1$ must be also
small: $|\lambda_1|^2\sim (Y_\nu Y_\nu^\dagger)_{11}$. However the
decay of the heavier (s)neutrinos does not need to be out of
equilibrium, so that $\lambda_2\sim1$ is allowed. In this range of
parameters,
\begin{equation}
  \label{eq:vv}
  \epsilon_{N_1}+\epsilon_{\tilde{N}_1}\approx 10^{-6}
\sqrt{\frac{M_1}{10^{6} \ {\rm GeV}}} \lambda_2 \sin \phi
\end{equation}
where $\phi $ is the relevant CP-violating phase which can be of
order of 1.

Now, let us discuss the wash-out processes. The evolution of the
numbers  of the relevant particles  is given by the following
Boltzman equations:
\begin{equation}
\label{nn1} \frac{d N_{N_1}}{dz}=-(D+S)(N_{N_1}-N_{N_1}^{eq}),
\end{equation}
and
\begin{equation}
\label{nb-l} \frac{d N_{B-L}}{dz}=-\epsilon_{N_1} D
(N_{N_1}-N_{N_1}^{eq})-W N_{B-L},
\end{equation}
where $N_{N_1}$ is the number of $N_1$ plus that of its
superpartner and $N_{B-L}$  denotes the baryon number minus the
number of standard model leptons [not including the right-handed
(s)neutrinos].
 In the above equations,
$z=M_1/T$ and $D$ and $S$ respectively represent the rates of the
decay and scattering of $N_1$ and $\tilde{N}_1^{(*)}$
($D=\Gamma_D/Hz$ and $S=\Gamma_S/Hz$).  $W \equiv \Gamma_W/Hz$ in
Eq.~(\ref{nb-l}) represents the rate of processes that erase the
produced $B-L$. Here,
 Since the rates of interactions of
$N_1$ and its superpartner are the same, it is not necessary to
consider the evolution of the number of $N_1$ and its
superpartner, separately \cite{dibari}. Moreover, writing
(\ref{nb-l}), we have used
$\epsilon_{N_1}=\epsilon_{\tilde{N}_1}$.

In the R-parity conserving case, it is shown that the dependence of
$S$ and $D$ on the seesaw parameters is through the combination
$$\tilde{m}_1= {(Y_\nu Y_\nu^\dagger)_{11} v^2 \over M_1}.$$
In the presence of the new interaction, there are new diagrams
contributing to both decay and scattering of right-handed
(s)neutrinos and the definition of $\tilde{m}_1$ has to be
modified to $$\tilde{m}_1={(Y_\nu
Y_\nu^\dagger)_{11}+|\lambda_1|^2 \over M_1}v^2.$$

We can divide the processes that contribute to $W$ into three
categories: i) Inverse decay processes of type $\ell H_u \to N_1$
and scattering such as $N_1 \ell \to \bar{t} q$, that involve
$N_1$ or $\tilde{N}_1$ along with a standard model lepton. The
dependence of the rate of these processes on seesaw parameters is
through the combination $(Y_\nu Y_\nu^\dagger)_{11}/M_1$; ii)
R-parity conserving $\Delta L=2$ processes such as $\ell \ell \to
H_u H_u$. The rate of these processes ($W^R$) can be written as
$$W^R=a M_1 \left(Y_\nu^\dagger \frac{1}{M} Y_\nu\right)^2.$$
Here, $a$ does not depend on the seesaw parameters. iii) Turning
on the R-parity violating couplings, new processes will take place
that contribute to wash-out of the produced $B-L$. The new
processes are of type $\ell H\to HH$ where $H$ collectively
denotes $H_u$, $H_d$ and their superpartners. Such processes take
place through virtual $N_2$ or $\tilde{N_2}$ exchange and their
rate can be estimated as
$$\sim a M_1\left|\lambda_2 (Y_\nu)_{2i}^*/M_2\right|^2.$$
Summing up the above discussion and remembering $|\lambda_1|^2\sim
(Y_\nu Y_\nu^\dagger)_{11}$, we conclude that, for a given value
of $\tilde{m}_1$, the effect of the wash-out processes in our
scenario is similar to the R-parity conserving case provided that
we replace $M_1$ with $M_1[\lambda_2/(Y_\nu)_{2i}]^2$. As a
result,  the wash-out factor for $\lambda_1\sim (Y_\nu)_{1i}$,
$\lambda_2\sim 1$ and $M_1\sim 10^6$ GeV (which results in
$\epsilon_{N_1} \sim 10^{-6}$) will be of order of  the wash-out
factor for $\lambda_2=0$ and $M_1\sim 10^{15}$ GeV. The wash-out
factor for the latter case is known. In \cite{Buchmuller:2002rq},
it is shown that for $\lambda_i=0$, $M_1\sim10^{15}$ GeV and
$\tilde{m}_1\sim 10^{-5}$ eV, an asymmetry ($\epsilon_{N_1}$) of
$10^{-6}$ is enough to explain the baryon asymmetry of the
universe provided that the initial number of the lightest
right-handed neutrino is thermal. Equivalently, we conclude that
for $\lambda_2\sim 1$, $|\lambda_1|^2\sim (Y_\nu
Y_\nu^\dagger)_{11}$ and $M_1\sim 10^6$ GeV, there is a range of
parameters [corresponding to $v^2\left(|\lambda_1|^2 +(Y_\nu
Y_\nu^\dagger)_{11}\right)/M_1\sim 10^{-5}$ eV] for which, through
the scenario discussed in this letter, the observed baryon
asymmetry of the universe can be produced. Notice that unlike the
case $\lambda_i=0$ and $M_1 =10^{15}$ GeV, the thermal production
of the right-handed (s)neutrinos in our scenario  can be realized
with safely low reheating temperature.

There is a subtlety here that should be noticed. As shown in
\cite{Buchmuller:2002rq}, for small values of $(Y_\nu)_{1i}$ which
correspond to $\tilde{m}_1\stackrel{<}{\sim}10^{-5}$~eV, even if the
reheating temperature is above $M_1$, the rates of interactions that
involve $Y_\nu$ will not be high enough to give rise to a thermal
initial number of $N_1$. In order to have a thermal initial number of
$N_1$ and $\tilde{N}_1$, there has to be ``another" mechanism for
production of right-handed (s)neutrinos.  In our scenario, there is a
natural mechanism to create initial thermal distribution which we
briefly discuss below.  Since in our scenario $\lambda_2\sim 1$, $N_2$
and $\tilde{N}_2$ maintain their thermal equilibrium and consequently
for temperatures below $M_1$ their numbers are negligible. As a result
an interaction of type
\begin{equation}
W^{N^3}=\lambda_{221}^{N^3}\hat{N}_2 \hat{N}_2 \hat{N}_1 \label{wn3}
\end{equation}
cannot contribute to the wash-out. However, at $T\stackrel
{>}{\sim} M_2$ (remember that we have assumed that $|M_2-M_1|/M_1
\sim 1$ so to have $T\sim M_2$ the reheating temperature does not
need to be far higher than $M_1$) the term in (\ref{wn3}) can
contribute to the  production of $N_1$ and $\tilde{N}_1$, giving
rise to a thermal distribution of $N_1$ and $\tilde{N}_1$ at
$T>M_1$. As a result, in this framework assuming a thermal initial
number of $N_1$ and $\tilde{N}_1$, even for very small values of
$(Y_\nu)_{1i}$ \cite{Buchmuller:2002rq}, is reasonable.

Determining the exact range of allowed parameters requires
detailed numerical calculation of the wash-out effects which is
beyond the scope of this letter and will be presented elsewhere.

Before concluding we note that for $\lambda_2 \sim 1$ the new term can
significantly affect the renormalization group equations of the Higgs
sector which may have consequences for radiative electroweak symmetry
breaking. In addition, the new term also slightly shifts the vacuum
expectation values of the scalars of the theory. The
$\tilde{N}$-dependant part of the scalar potential is
\begin{eqnarray}& & \sum_i|M_i
  \tilde{N}_i+(Y_\nu)_{ij}\tilde{L}_jH_u+\lambda_i H_d H_u + \sum_{jk}
  \lambda^{N^3}_{ijk} \tilde{N}_j\tilde{N}_k|^2+ \cr & &
  \sum_{ij}\left[ m_0^2|\tilde{N_i}|^2+B_\nu M_i(\tilde{N}_i^2+{\rm
      H.c.})\right. \cr & &+\left. [A_\lambda^i \tilde{N}_i H_d H_u
    +(A_\nu)_{ij}\tilde{N}_i \tilde{L}_j H_u + {\rm H. c.}]\right]
\end{eqnarray}
Because of the new term the right-handed sneutrinos develop very small
vacuum expectation values:
$$
\left| \vev{ \tilde{N}_i}\right| \simeq {\lambda_i v^2 \sin\beta
  \cos\beta\over M_i}\ll v.$$
Expanding the superpotential around $\vev{\tilde{N}_i}$ we obtain
a tiny correction to the $\mu$ term. Moreover, we obtain the
following bilinear R-parity violating term \begin{equation}
\label{vev} \sum_i (Y_\nu)_{ij}\epsilon_{\alpha \beta} {\lambda_i
v^2 \sin
  \beta \cos \beta \over M_i}\hat{L}_j^\alpha \hat{H}_u^\beta.\end{equation}

As discussed in \cite{Berezinsky:1991sp}, such term gives rise to the
decay of lightest neutralino. For the specific parameter range studied
in this letter, we can make the following estimate
 \begin{eqnarray} & &\Gamma (\chi \to \nu_i+ e^-+ e^+)\sim 10 ~ {\rm
  sec}^{-1}~ \lambda_2^2\times \cr & &
  {\cos^2 \beta \over 0.01}{{\rm Max}[(Y_\nu)_{2i}^2,(Y_\nu)_{2e}^2] \over 10^{-9}} \left({10^6 {\rm GeV}\over M_2}
  \right)^2 \left( {m_\chi \over 100 {\rm GeV}}\right)^3 \nonumber .
\end{eqnarray}
This implies that, if produced at colliders, like the Large Hadron
Collider, the lightest neutralino would leave the detector before
decaying, leading to the same missing energy signature as in the
minimal supersymmetric standard model. However, in our model the
lightest neutralino typically decays before the epoch of
nucleosynthesis. Thus it cannot serve as  dark matter, which needs
another candidate, like the axion.  The neutrino mass induced by
Eq.~(\ref{vev}) will be of order $(Y_\nu \lambda v^2 \sin\beta
\cos\beta/M)^2/m_{susy}$\cite{Hirsch:2004he}, completely
negligible in comparison with the seesaw effect, $Y_\nu^2
v^2\sin^2\beta/M$.


In conclusion we have suggested a simple variant of the supersymmetric
seesaw mechanism where the thermal leptogenesis is assisted by an
explicit R-parity violating term involving the heavy right-handed
neutrino supermultiplets, $N_i$.  In this scenario, the lightest
right-handed neutrino ($N_1$) and its superpartner ($\tilde{N}_1$)
could be as light as $\sim10^6$~GeV or less and still account for the
baryon asymmetry of the universe, avoiding overproduction of
gravitinos which plagues the R-parity conserving thermal leptogenesis.

This work was supported by Spanish grant BFM2002-00345 and by the EC
RTN network MRTN-CT-2004-503369.  We thank A.  Yu. Smirnov for
encouragement.



\end{document}